\documentclass[a4paper,12pt,reqno]{amsart}
\pdfoutput=1

\usepackage{a4}
\usepackage{amsthm}
\usepackage[T1]{fontenc}
\usepackage[utf8]{inputenc}
\usepackage[british]{babel}
\usepackage{todonotes}
\usepackage{scalerel}
\usepackage[unicode,psdextra]{hyperref}
\usepackage{tikz}
\usetikzlibrary{decorations.pathmorphing,shapes}
\usetikzlibrary{arrows.meta,calc}
\def\centerarc[#1](#2)(#3:#4:#5)% Syntax: [draw options] (center) (initial angle:final angle:radius)
    { \draw[#1] ($(#2)+({#5*cos(#3)},{#5*sin(#3)})$) arc (#3:#4:#5); }

\usepackage{scrlayer-scrpage}
\newtheorem{definition}{Definition}

\newtheorem{theorem}[definition]{Theorem}

\newtheorem{corollary}[definition]{Corollary}

\DeclareMathOperator{\Res}{Res}

\makeatletter
\@addtoreset{definition}{section}
\@addtoreset{equation}{section}
\makeatother

\makeatletter
\let\@wraptoccontribs\wraptoccontribs
\makeatother

\allowdisplaybreaks[4]

\begin{document}

\title[An irregular spectral curve for the generation of bipartite maps in topological recursion]{An irregular spectral curve for the generation of bipartite maps in topological recursion}

\author[J. Branahl]{Johannes Branahl\textsuperscript{1}\href{https://orcid.org/0000-0002-4107-5039}{\scaleto{\includegraphics{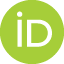}}{10pt}}}
 \author[A. Hock]{Alexander Hock\textsuperscript{2} \href{https://orcid.org/0000-0002-8404-4056}{\scaleto{\includegraphics{ORCID-iD_icon-64x64.png}}{10pt}}}
 
\address{\textsuperscript{1}Mathematisches Institut der
  Westfälischen Wilhelms-Universit\"at \hfill \newline
Einsteinstr.\ 62, 48149 M\"unster, Germany \hfill \newline
{\itshape e-mail:} \normalfont
\texttt{j\_bran33@uni-muenster.de}}

\address{\textsuperscript{2}Mathematical Institute, University of Oxford,
  \newline
  Andrew Wiles Building, Woodstock Road, OX2 6GG, Oxford, United Kingdom
  \newline
  {\itshape e-mail:} \normalfont \texttt{alexander.hock@maths.ox.ac.uk}}

\begin{abstract}
  We derive an efficient way to obtain generating functions of bipartite maps of arbitrary genus and boundary length using a spectral curve as initial data for the framework of topological recursion. Based on an earlier result of Chapuy and Fang counting these maps and having a structural proximity to topological recursion, we deduce the corresponding spectral curve which has a strong relation to the spectral curve giving rise to generating functions of ordinary maps. In contrast to ordinary maps, the spectral curve is an irregular one in the sense of Do and Norbury. It generalises the irregular curve for the enumeration of Grothendieck's dessins d'enfant.
\end{abstract}


\maketitle
\markboth{\hfill\textsc\shortauthors}{\textsc{{An irregular spectral curve for the generation of bipartite maps in topological recursion}\hfill}}

\section{Introduction and Main Result}
 The enumeration of maps has a long history, in which the techniques and tools became more and more efficient and the classes of maps more and more sophisticated: In his \textit{Census of Planar Maps}, William Tutte achieved groundbreaking progress in the 1960's \cite{Tuttbij}.  Bender and Canfield then left the realm of planar maps in the 1980's and also took an embedding into higher-genus surfaces into consideration \cite{Bender}. In the 2000's, the branch of mathematical physics established a powerful and efficient universal procedure to reach all topological sectors in a recursive way: Topological recursion (TR) of Chekhov, Eynard and Orantin \cite{Eynard:2007kz,Chekhov:2006vd} built a bridge between enumerative and complex geometry (and, based on the work \cite{Kontsevich:1992ti}, bridges to intersection theory and integrable hierachies, which we will neglect here) and thus covered numerous, seemingly disconnected areas of mathematical fields, by one universal recursion procedure.

Topological recursion possesses the initial data $(\Sigma,x,y,B)$, where $x:\Sigma\to \Sigma_0$ is a ramified covering of Riemann surfaces, $\omega_{0,1}=y\, dx$ is a meromorphic differential 1-form on $\Sigma$ regular at the ramification points and $\omega_{0,2}=B$ a symmetric bilinear differential form on $\Sigma\times \Sigma$ with double pole on the diagonal and no residue. From this initial data, TR computes recursively in the negative Euler characteristic $-\chi=2g+n-2$ an infinite sequence of symmetric meromorphic $n$-forms $\omega_{g,n}$ on $\Sigma^n$ with poles only at the ramification points for $-\chi>0$. The precise formula and more details are given in Ch. \ref{ch:proof}. For specific choices of the initial data $(\Sigma,x,y,B)$, the meromorphic $n$-forms are encoding some enumerative problems.

The prime example of this framework was the recursive computation of generating functions counting objects known in the literature as \textit{ordinary maps}  (a very readable derivation can be found in \cite{Eynard:2016yaa}):
\begin{theorem}[\cite{Eynard:2016yaa}]
\label{th:eyn}
The spectral curve $(\overline{\mathbb{C}},x_{ord},y_{ord}, \frac{dz_1\, dz_2}{(z_1-z_2)^2})$ with
\begin{align*}
	x_{ord}(z)=\gamma \bigg ( z+ \frac{1}{z} \biggl ) \qquad y_{ord}(z)= \sum_{k=0}^{d-1} u_{2k+1}z^{2k+1}
\end{align*}
where
\begin{align*}
	\gamma^2=1+\sum_{k\geq 1} t_{2k} \binom{2k-1}{k} \gamma^{2k}, \qquad u_{2k+1}=\gamma \bigg(\delta_{k,0}-\sum_{j\geq k+1}t_{2j}\binom{2j-1}{j+k}\gamma^{2j-2}\bigg)
\end{align*}
computes via TR (see formula \eqref{BTR-intro}) generating functions for the enumeration of ordinary maps with $n$ marked faces of even boundary lengths. The faces have even degrees up to $2d$, where a face of degree $2k$ is weighted by $t_{2k}$.
\end{theorem}\noindent
The theorem includes in general also faces of odd degree, but for later purposes, we want to state it in this form.

Several more classes of maps, e.g. subsets of the ordinary maps, were then discovered to be governed by TR, as \textit{ciliated} and \textit{fully simple maps} \cite{Borot:2017agy,Borot:2021eif}.

  In this letter, we will focus on another subset of maps, the \textit{bipartite maps} containing only those ordinary maps of even face degrees, for which the corresponding maps have vertices in black and white such that no monochromatic edge occurs. A bipartite map is called \textit{rooted}, if one edge is distinguished and oriented. This \textit{rooted edge} (also called marked edge) conventionally  has its origin in a white vertex (the \textit{root vertex}). Rooting an edge creates a boundary of a certain even length $2l_k$ following the face to the right of the rooted edge. Several edges can be rooted such that the roots do not correspond to the same boundary. Bipartite maps already showed up in the context of TR, namely in \cite{Chapuy2016} in which the authors were motivated by TR and established a recursive formula sharing many characteristics with the TR \footnote{A more formal, but less illustrative definition for bipartite maps can be found in \cite{Chapuy2016}}. However, the aim of their work is rather to prove rationality statements about bipartite maps and is thus written more in a combinatorist's language. All these statements are a direct consequence of TR. Their recursion and its proof are mainly built on ideas of TR, but no spectral curve was provided. Thus, the relation of their work to complex geometry will be established in the following Chapter \ref{ch:proof}. We will deduce:
\begin{theorem}\label{th:main}
The spectral curve $(\overline{\mathbb{C}},x_{bip},y_{bip},\frac{dz_1\,dz_2}{(z_1-z_2)^2})$ with 
\begin{align*}
	x_{bip}(z)=\gamma^2\bigg ( z+ \frac{1}{z} \bigg ) + 2\gamma^2 \qquad y_{bip}(z)=\frac{\sum_{k=0}^{d-1} u_{2k+1}z^{k+1}}{\gamma (1+z)}
\end{align*}
computes via TR generating functions for the enumeration of bipartite maps with $n$ marked faces (or rooted edges) of even boundary lengths. The faces have even degrees up to $2d$, where a face of degree $2k$ is weighted by $t_{2k}$. We have the following relation to Thm. \ref{th:eyn}:
\begin{align*}
x_{bip}(z^2)=x_{ord}(z)^2 \qquad \quad y_{bip}(z^2) = \frac{y_{ord}(z)}{ x_{ord}(z)}
\end{align*}
\end{theorem}
In order to avoid misunderstandings, we would like to mention that an unconventional definition of bipartite maps, deviating from the one in this paper, is given in \cite{Eynard:2016yaa} and coincides just for genus zero and one boundary. 

Given these two spectral curves for ordinary and bipartite maps, the machinery of TR gives rise to generating functions as follows: Let $\tilde{\mathcal{T}}^{(g)}_{2l_1,...,2l_n}$ denote the generating function of bipartite maps with a natural embedding into a genus-$g$ surface with $n$ boundaries of length $2l_1,...,2l_n$ ($n$-fold rooted bipartite maps) and in the same manner $\mathcal{T}^{(g)}_{2l_1,...,2l_n}$ for ordinary maps with faces of even degree. Note that in particular $2^{n-1}\tilde{\mathcal{T}}^{(0)}_{2l_1,...,2l_n}=\mathcal{T}^{(0)}_{2l_1,...,2l_n}$ holds for genus $g=0$, however not for $g>0$, where only a small subset of ordinary maps are still bipartite. The prefactor $2^{n-1}$ has an easy combinatorial explanation: As described earlier, the \textit{root vertex} is by convention a white one, the black-white colouring of the vertices is completely determined by fixing a root. Ignoring  the colouring, as it is done for ordinary maps, another boundary can have twice the number of labellings. Inductively, this gives rise to $2^{n-1}$ distinct graphs, if $n$ faces are marked, as ordinary maps.

 Define the \textit{correlators} $W$ and $\tilde W$  as 
  \begin{align}
\label{resolv}
 W_n^{(g)}(x_{ord,1},...,x_{ord,n}) =   \sum_{l_1,...,l_n=1}^\infty \frac{\mathcal{T}^{(g)}_{2l_1,...,2l_n}}{x_{ord,1}^{2l_1+1}...x_{ord.n}^{2l_n+1}}\nonumber \\
 \tilde {W}_n^{(g)}(x_{bip,1},...,x_{bip,n}) =   \sum_{l_1,...,l_n=1}^\infty \frac{\tilde{\mathcal{T}}^{(g)}_{2l_1,...,2l_n}}{x_{bip,1}^{l_1+1}...x_{bip,n}^{l_n+1}}
\end{align}
from which the generating functions can be read off as a simple residue operation, e.g. for bipartite maps:
\begin{align*}
\tilde{\mathcal{T}}^{(g)}_{2l_1,...,2l_n} = (-1)^n \Res_{x_{bip,1}...x_{bip,n}\to \infty}  x_{bip}^{l_1}\cdot ...\cdot x_{bip}^{l_n} \tilde {W}_n^{(g)}(x_{bip,1},...,x_{bip,n})dx_{bip,1}...dx_{bip,n}
\end{align*}
The crucial connection to the infinite sequence of meromorphic $n$-forms $\omega_{g,n}$ generated by TR is the following identification for $2g+n-2>0$
\begin{align}
	\omega_{g,n}(z_1,...,z_n)=\tilde{W}^{(g)}_n(x_{bip}(z_1),...,x_{bip}(z_n))dx_{bip}(z_1)...dx_{bip}(z_n).
\end{align}
For the stable topologies $2g+n-2\leq 0$, the situation is a bit subtle.

From Thm. \ref{th:main}, we deduce the following equivalent representation of the generating functions of bipartite maps, building the bridge to TR:
\begin{corollary}
\label{cor1}
Let $\omega_{g,n}$ be the correlators of TR  generated by \eqref{BTR-intro} with $(x_{bip},y_{bip})$ given by Theorem \ref{th:main}. Then  $\tilde{\mathcal{T}}^{(g)}_{2l_1,...,2l_n}$ can be achieved as follows:
\begin{align*}
\tilde{\mathcal{T}}^{(g)}_{2l_1,...,2l_n} = (-1)^n \Res_{z_1,...,z_n \to \infty} x_{bip}(z_1)^{l_1} \cdot ... \cdot x_{bip}(z_n)^{l_n}  \omega_{g,n}(z_1,...,z_n)
\end{align*}
\end{corollary}
Analogously, generating functions for ordinary maps are obtained from the spectral curve of Theorem \ref{th:eyn} (see \cite{Eynard:2016yaa} for more details). This spectral curve, together with the recursion formula of \cite[Thm. 3.9]{Chapuy2016}, will be the basis for the proof of Theorem \ref{th:main} by direct identification.

\section{Proof}
\label{ch:proof}
\subsection{Reminder of Previous Results}
First, we briefly recapitulate the procedure of topological recursion. Starting with the initial data, the spectral curve  $(\Sigma,x,y,B)$, TR constructs recursively in $2g+n-2$ an infinite sequence of meromorphic $n$-forms  $\omega_{g,n}$, starting with
  \begin{align*}
\omega_{0,1}(z) = y(z)\, dx (z) \qquad \omega_{0,2}(z_1,z_2) = B(z_1,z_2),
\end{align*}
 via the following residue formula:
\begin{align}
&  \omega_{g,n+1}(I,z)
  \label{BTR-intro}
  \\
  & =\sum_{\beta_i}
  \Res\displaylimits_{q\to \beta_i}
  K_i(z,q)\bigg(
  \omega_{g-1,n+2}(I, q,\sigma_i(q))
  +\hspace*{-1cm} \sum_{\substack{g_1+g_2=g\\ I_1\uplus I_2=I\\
            (g_1,I_1)\neq (0,\emptyset)\neq (g_2,I_2)}}
  \hspace*{-1.1cm} \omega_{g_1,|I_1|+1}(I_1,q)
  \omega_{g_2,|I_2|+1}(I_2,\sigma_i(q))\!\bigg).
\end{align}
Here $I=\{z_1,\dots,z_n\}$ is a collection of $n$ variables $z_j$, the sum is over the ramification points
$\beta_i$ of $x$ defined by $dx(\beta_i)=0$. The kernel $K_i(z,q)$ is
defined in the vicinity of $\beta_i$ by
$K_i(z,q)=\frac{\frac{1}{2}\int^{q}_{\sigma_i(q)}
  B(z,q')}{\omega_{0,1}(q)-\omega_{0,1}(\sigma_i(q))}$, where
$\sigma_i\neq \mathrm{id}$ is the local Galois involution
$x(q)=x(\sigma_i(q))$ near $\beta_i$, and $\beta_i$ as a fixed point.

A TR-like formula to recursively generate correlators for bipartite maps was found in the aforementioned paper \footnote{We adapt the notation of \cite{Chapuy2016} to the TR literature by $p_k \mapsto t_{2k}$, $z \mapsto \gamma^2$, $u \mapsto z$, $F_g\mapsto U_g$}:
\begin{theorem}[\cite{Chapuy2016}]
\label{th:chap}
Let  $x(z)= \frac{z}{(1+z\gamma^2)^2}$. A correlator $U_g(x(z))= \sum_{l=1}^\infty \tilde{\mathcal{T}}^{(g)}_{2l} x^l$, $g>1$, can be recursively obtained in the following way:
\begin{align*}
U_g(x(z)) = \frac{1}{P(z)} \Res_{q \to \pm \frac{1}{\gamma^2}} \frac{P(q)}{z-q} \frac{x(q)}{Y(q)} \bigg (U_{g-1}^{(2)}(q)+ \sum_{\substack{g_1+g_2=g\\g_i>0}} U_{g_1}(q)U_{g_2}(q) \bigg ) 
\end{align*}
with $P(q)=\frac{1-\gamma^2 q}{1+ \gamma^2 q}$, $Y(q)$ see below. $U^{(2)}_g=\sum_{l_1,l_2=1}^\infty \tilde{\mathcal{T}}^{(g)}_{2l_1,2l_2} x^{l_1+l_2}$.
\end{theorem}

\subsection{Proof of the Spectral Curve}
We remark that most of the work was already done by Chapuy and Fang \cite{Chapuy2016} by proving Thm. \ref{th:chap}. However, the authors omitted the decisive step to read off a spectral curve from their TR-like formula. This shall be done in the following, making Thm. \ref{th:main} rather to a corollary of Thm. \ref{th:chap}. However, analysing the somewhat unusual geometry of $(x_{bip},y_{bip})$ and its deep relation to $(x_{ord},y_{ord})$ makes it worth to treat the derivation of the spectral curve in this article.
The deduction of $(x_{bip},y_{bip})$ that finally turns Thm. \ref{th:chap} into topological recursion works as follows: 
\begin{itemize}
 \item $x_{bip}$: The work of Chapuy and Fang mainly relies on two important variable transformations. The first is the definition of $\gamma^2$, arising already for ordinary maps and earlier works of Bender and Canfield \cite{Bender}. The second, $x(z)= \frac{z}{(1+z\gamma^2)^2}$ will determine $x_{bip}$. Thm. \ref{th:chap} creates generating functions as a series in positive powers of $x$. Sending $z \to \frac{z}{\gamma^2}$ and then taking the reciprocal of $x$ gives the correct curve ramified covering.  We confirm this with the relation $x_{bip}(z^2)=x_{ord}^2(z)$ together with a comparison of the correlators $W$ and $\tilde W$, up to a global factor of $\frac{1}{x_{bip}}$ on which we comment later - this factor becomes decisive for the geometry of the spectral curve.
\item $y_{bip}$: Analogously to ordinary maps, the expression $y_{bip}(z)-y_{bip}(\sigma(z))$ can be directly read off from the kernel representation of the Tutte equation for the disk. This kernel $Y(z)=y_{bip}(1/z)-y_{bip}(z)$ is already given in Prop. 3.3 in \cite{Chapuy2016} and shows up in the recursion formula Thm. \ref{th:chap} as well. After changing the variables as for $x_{bip}$, we can extract from \cite[Chap. 5.1]{Chapuy2016} a suitable expression for $Y(z)\cdot x_{bip}(z)$:  
\begin{align*}
&\qquad \quad Y(z)\cdot x_{bip}(z) =\\
& \qquad \quad  \gamma^2\frac{(1+z)^2}{z} - (1+z)\bigg [2- \sum_{k=1}^d t_{2k} \gamma^{2k} \bigg (\sum_{l=1}^{k-1}z^l \binom{2k-1}{k+l} - \sum_{l=-k}^0 z^l \binom{2k-1}{k+l} \bigg ) \bigg ].
\end{align*}
Inserting the implicit equation of $\gamma^2$ from Theorem \ref{th:eyn} in the first term cancels partially the terms for $l=-1,0$. After some further lengthy but trivial algebra, the expression can be ordered in positive and negative powers of $z$, where the positive powers give  
\begin{align}\label{yeq}
	y_{bip}(z)\cdot x_{bip}(z)= 1+z-(1+z)\sum_{k=1}^{d-1} \sum_{j\geq k+1} t_{2j} \binom{2j-1}{j+k} \gamma^{2j} z^k.
\end{align}
Finally, the definition of $u_{2k+1}$ in terms of $t_{2j}$ yields the desired identifications shown in Thm. \ref{th:main}. $ y_{bip}(z^2) = \frac{y_{ord}(z)}{ x_{ord}(z)}$ follows immediately. \qed
\end{itemize}

We emphasise that this result does not only clarify the complex geometric viewpoint on the enumeration of bipartite maps, but also extends the result of \cite{Chapuy2016} to an arbitrary number $n$ of boundaries, whereas the construction of Thm. \ref{th:chap} only gives rise to $\omega_{g,1}$ and hence $\tilde{\mathcal{T}}^{(g)}_{2l}$ instead of the more general result being $\tilde{\mathcal{T}}^{(g)}_{2l_1,...,2l_n}$ (compare with eq.~\ref{resolv}).

\section{Discussion of the Result}
\label{ch:disc}

\subsection{On the Irregularity}
Of particular interest is the special geometry of the spectral curve for bipartite maps. Its branch cut goes from $x_{bip}(1)=a=4\gamma^2$ to $x_{bip}(-1)=b=0$. We naturally have the same Zhukovsky parametrisation as for $x_{ord}(z)$:
\begin{align}\label{zhu}
\frac{a+b}{2}+\frac{a-b}{4} \bigg (z + \frac{1}{z} \bigg ) \qquad  \mathrm{and} \qquad \sqrt{(x-a)(x-b)} =\gamma^2\bigg (z - \frac{1}{z} \bigg )
\end{align}
However, the branch point at $0=x(\beta_2)$ corresponding to the ramification point $\beta_2=-1$ affects the pole structure of all $\omega_{g,n}$ - the highest degree of the poles is different for the two ramification points. Due to the fact that $y_{bip}$ is not regular at the ramification point $\beta_2=-1$ (whereas as required $\omega_{0,1}=y \,dx$ is still regular), the maximum order of poles $\frac{1}{(z+1)^k}$ is reduced in comparison to the poles $\frac{1}{(z-1)^l}$. However, this does not change the fact that one can generate symmetric $n$-forms from $(x_{bip},y_{bip})$.   Despite the uncommon pole distribution at the ramification points, the universal symmetry under the Galois involution naturally holds: 
 \begin{align*} 
\frac{\omega_{g,n}(z,z_I)}{dx(z)} + \frac{\omega_{g,n}(\frac{1}{z},z_I)}{dx(\frac{1}{z})}=0 \quad \forall 2g+n-2>0 
\end{align*}
There exists a more detailed analysis of spectral curves in which the meromorphic function $y(z)$ has a simple pole at a $z=\beta_j$ being a root of $dx(z)$. Do and Norbury named this class of spectral curves \textit{irregular spectral curves} \cite{Norbury2014}. Remarkably, the maximum order of the poles in $\omega_{g,n}$ at $z=\beta_j$ reduces generally to $2g$ (independent of $n$), in contrast to $6g-4+2n$ for regular curves. The class of irregular curves has only a handful of concrete meaningful examples so far, to which we add a generalisation. We want to mention the somewhat artificially constructed analogue of the Airy curve $x=y^2$, given by $xy^2=1$, treated in \cite{Norbury2016} where it is named \textit{Bessel curve}, and the more complicated curve $xy^2-xy=-1$ giving rise to a weighted count of Grothendieck's \textit{dessins d'enfant}, being bicoloured graphs embedded into a connected orientable surface, such that the complement
is a union of disks. In other words, the underlying graphs of dessins d'enfant are bipartite. In case of exactly three branching points on the projective line, the generating functions for their enumeration corresponds to complex matrix models \cite{ambj}, as it is the case for the enumeration of bipartite maps of arbitrary topology (complex 1-matrix model). Therefore, it is not surprising that the parametric representation of  $xy^2-xy=-1$, given by 
 \begin{align*} 
x(z)=z+ \frac{1}{z}+2 \qquad y(z) = \frac{z}{1+z} \; ,
\end{align*}
is the special case of $(x_{bip},y_{bip})$ when setting $t_{2k}=0 \, \forall k$, being the complex 1-matrix model with Gaussian potential and implying $u_{2k+1}=0$ $\forall k>0$ and $u_1=\gamma^2=1$. As the weights $t_k$ keep track of the unmarked faces, this special case corresponds to bipartite maps with all faces marked. The proximity to the curve  $y^2-xy=-1$, solving the hermitian 1-matrix model with only Gaussian potential (widely studied in the literature), supports the more general relation between the hermitian and complex matrix model in the parametric representation of the two curves
\begin{align}
x_{bip}(z^2)=x_{ord}(z)^2 \qquad \quad y_{bip}(z^2) = \frac{y_{ord}(z)}{ x_{ord}(z)} \; .
\end{align}
  For illustrative purposes, we give $\omega_{1,1}$ arising from an irregular curve and enumerating toroidal bipartite maps with a single boundary, as an example and set for brevity $\tilde{y}_{bip}(z) = \frac{1}{\gamma} \sum_{k=0}^{d-1}u_{2k+1} z^{k+1}$
 \begin{align*} 
\omega_{1,1}(z) =& \frac{1}{16 \gamma^2 (1+z)^2 \tilde{y}'_{bip}(-1)}- \frac{1}{16 \gamma^2 (z-1)^4 y'_{bip}(1)}\\
&-\frac{1}{16 \gamma^2 (z-1)^3 y'_{bip}(1)} + \frac{3 y'_{bip}(1)+3 y''_{bip}(1)+ y'''_{bip}(1)}{96 \gamma^2 (z-1)^2 y'^2_{bip}(1)}
\end{align*}
This corresponds to the maximum order of poles of $2g$ at $z=-1$. \\ \\
Finally, we want to collect some interesting  open questions:  
It is known \cite{Borot:2017agy,Borot:2021eif} that the exchange of $x_{ord}$ and $y_{ord}$ gives rise to generating functions of fully simple maps. Does any sort of exchange of $x_{bip}$ and $y_{bip}$ have a comparable strong implication? The general behaviour of correlators in TR under exchange of $x$ and $y$ was e.g. considered in \cite{Borot:2021,Hock:2022wer}. Another question arises from the matrix models as realisations of those various types of maps. As known from the classical literature, bipartite maps arise from the complex matrix model, having a structural equivalence to the hermitian 2-matrix model \cite{Eynard:2005}. This model is (for certain boundary structures) already solved by TR. What is the relation between the two distinct spectral curves?

 A final question is dedicated only to quadrangulations. In \cite{Branahl:2020yru} the quartic Kontsevich model (QKM) was shown to be solvable in terms of correlators $\omega_{g,n}$ that follow an extension of TR. In this so-called blobbed topological recursion (BTR; general framework developed in \cite{Borot:2015hna}), the $\omega_{g,n}$ split into parts with poles at the ramification points (polar part) and with poles somewhere else (holomorphic part). In \cite{Branahl:2021} it was stated that the $\omega_{g,n}$ of BTR in the QKM are generating functions for ordinary (rooted) quarangulations, whereas according to \cite{Branahl:2022} the complex analogue of this model (namely, the LSZ model with quartic potential) follows topological recursion only and gives rise to generating functions of bipartite (rooted) quadrangulations. All that is reached in the combinatorial limit of these two matrix models with external fields, in which the external matrix has an $N$-fold degenerate eigenvalue and is thus a multiple to the identity matrix. This basically gives the same partition functions as in the 1-matrix models. Understanding this different approach to the partition functions of the hermitian and complex 1-matrix model from the beginning will be an interesting challenge for the future.

\subsection{Example of Quadrangulations}\label{a}
 In order to underpin the correctness of our spectral curve, let us only allow for $t_4 \neq 0$ and $n=l=1$, yielding (with $u_1=\frac{1}{\gamma}$ and $u_3=-t_4\gamma^3$):
 \begin{align*} 
x_{bip}(z) = \gamma^2 \bigg( z+\frac{1}{z} \bigg) + 2\gamma^2, \qquad y_{bip}(z)= \frac{\frac{z}{\gamma^2}-t_4 \gamma^2 z^2}{1+z}, \qquad \gamma^2 = \frac{1-\sqrt{1-12t_4}}{6t_4}.
\end{align*}
 The expansions in $t_4$ by computer algebra can be found in Tab. \ref{tab1}. Bipartite rooted quadrangulations are in particular interesting, since Tutte's famous bijection \cite{Tuttbij} relates them to rooted ordinary maps for faces of any (not only even) degree.
\begin{table}[h]
\centerline{\begin{tabular}[h!b]{|c|c|c|c||c|c|c|}
\hline
Order &$\tilde{\mathcal{T}}_2^{(0)}$&$\tilde{\mathcal{T}}_2^{(1)}$&$\tilde{\mathcal{T}}_2^{(2)}$ &$  \mathcal{T}_2^{(0)}$&$ \mathcal{T}_2^{(1)}$&$ \mathcal{T}_2^{(2)}$ \\
\hline
$(t_4)^0$ & 1 &0 &0  & 1 & 0 & 0  \\
\hline
$(t_4)^1$ & 2 & 0 & 0 & 2 & 1 & 0  \\
\hline
$(t_4)^2$ & 9 & 1 & 0 & 9 & 15 & 45  \\
\hline
$(t_4)^3$ & 54 & 20 & 0 & 54 & 198 & 2007 \\
\hline
$(t_4)^4$ & 378 & 307 & 21 & 378  & 2511 & 56646 \\
\hline
$(t_4)^5$ & 2916  & 4280 & 966  & 2916  & 31266 & 1290087  \\
\hline
\end{tabular}}
\hspace*{1ex}
\caption{These numbers are generated by Thm \ref{th:main} and Thm. \ref{th:eyn} together with Cor. \ref{cor1} and coincide with \cite{Bender} and with OEIS no. \href{https://oeis.org/A006300}{A006300} $(g=1)$ and no. \href{https://oeis.org/A006301}{A006301} ($g=2$) for $\tilde{\mathcal{T}}_{2}^{(g)}$.}
\label{tab1}
\end{table}

\section*{ Acknowledgements}
We thank Guillaume Chapuy and Wenjie Fang for helpful discussions. JB is supported\footnote{``Funded by
  the Deutsche Forschungsgemeinschaft (DFG, German Research
  Foundation) -- Project-ID 427320536 -- SFB 1442, as well as under
  Germany's Excellence Strategy EXC 2044 390685587, Mathematics
  M\"unster: Dynamics -- Geometry -- Structure."} by the Cluster of
Excellence \emph{Mathematics M\"unster}. He would like to thank the University of Oxford for its hospitality. The work of JB at the University of Oxford was additionally financed by the \emph{RTG 2149 Strong and Weak
Interactions – from Hadrons to Dark Matter}. AH is supported by
the Walter-Benjamin fellowship\footnote{``Funded by
  the Deutsche Forschungsgemeinschaft (DFG, German Research
  Foundation) -- Project-ID 465029630}.
 
\section*{Data Availability}

Data sharing is not applicable to this article as no datasets were generated or analyzed
during the current study, except of the numbers given in  Tab. \ref{tab1}. 

\section*{Statements and Declarations}

\textbf{Conflict of interests/competing interests:}
On behalf of the authors, the corresponding author states that there is no conflict of
interest.
 
\bibliographystyle{halpha-abbrv}
\bibliography{bip_maps}

\begin{thebibliography}{BCGF21}
\expandafter\ifx\csname url\endcsname\relax
  \def\url#1{\texttt{#1}}\fi
\expandafter\ifx\csname doi\endcsname\relax
  \def\doi#1{\burlalt{doi:#1}{http://dx.doi.org/#1}}\fi
\expandafter\ifx\csname urlprefix\endcsname\relax\def\urlprefix{URL }\fi
\expandafter\ifx\csname href\endcsname\relax
  \def\href#1#2{#2}\fi
\expandafter\ifx\csname burlalt\endcsname\relax
  \def\burlalt#1#2{\href{#2}{#1}}\fi

\bibitem[1]{Tuttbij}
W.~Tutte.
\newblock {A Census of Planar Maps.}
\newblock {\em Canadian Journal of Mathematics}, 15:249--271, 1963.
\newblock \doi{10.4153/CJM-1963-029-x}.

\bibitem[2]{Bender}
E.~Bender and E.~R. Canfield.
\newblock {The Enumeration of Maps on the Torus and the Projective Plane}.
\newblock {\em Canadian Mathematical Bulletin}, 31(3):257--271, 1988.
\newblock \doi{10.4153/CMB-1988-039-4}.


\bibitem[3]{Eynard:2007kz}
B.~Eynard and N.~Orantin.
\newblock {Invariants of algebraic curves and topological expansion}.
\newblock {\em Commun. Num. Theor. Phys.}, 1:347--452, 2007,
  \burlalt{math-ph/0702045}{http://arxiv.org/abs/math-ph/0702045}.
\newblock \doi{10.4310/CNTP.2007.v1.n2.a4}.

 

\bibitem[4]{Chekhov:2006vd}
L.~Chekhov, B.~Eynard, and N.~Orantin.
\newblock {Free energy topological expansion for the 2-matrix model}.
\newblock {\em JHEP}, 12:053, 2006,
  \burlalt{math-ph/0603003}{http://arxiv.org/abs/math-ph/0603003}.
\newblock \doi{10.1088/1126-6708/2006/12/053}.

\bibitem[5]{Kontsevich:1992ti}
M.~Kontsevich.
\newblock {Intersection theory on the moduli space of curves and the matrix
  Airy function}.
\newblock {\em Commun. Math. Phys.}, 147:1--23, 1992.
\newblock \doi{10.1007/BF02099526}.
 
 
\bibitem[6]{Eynard:2016yaa}
B.~Eynard.
\newblock {\em {Counting Surfaces}}, volume~70 of {\em Prog. Math. Phys.}
\newblock Birkh{\"a}user/ Springer, 2016.
\newblock \doi{10.1007/978-3-7643-8797-6}.

 \bibitem[7]{Borot:2017agy}
G.~Borot and E.~Garcia-Failde.
\newblock {Simple maps, Hurwitz numbers, and Topological Recursion}.
\newblock {\em Commun. Math. Phys.}, 380(2):581--654, 2020,
  \burlalt{1710.07851}{http://arxiv.org/abs/1710.07851}.
\newblock \doi{10.1007/s00220-020-03867-1}.

\bibitem[8]{Borot:2021eif}
G.~Borot, S.~Charbonnier, and E.~Garcia-Failde.
\newblock {Topological recursion for fully simple maps from ciliated maps}.
\newblock 2021, \burlalt{2106.09002}{http://arxiv.org/abs/2106.09002}.

\bibitem[9]{Chapuy2016}
G.~Chapuy and W.~Fang.
\newblock {Generating functions of bipartite maps on orientable surfaces}.
\newblock {\em Electron. J. Combin.}, 23(3):37, 2016.
 \burlalt{1502.06239}{http://arxiv.org/abs/1502.06239}
\newblock \doi{10.37236/5511}.

\bibitem[10]{Norbury2014}
N.~Do and P.~Norbury.
\newblock {Topological recursion for irregular spectral curves}.
\newblock {\em Journal of the London Mathematical Society}, 97, 2018.


\bibitem[11]{Norbury2016}
N.~Do and P.~Norbury.
\newblock {Topological recursion of the Bessel curve}. 
\newblock 2016, \burlalt{1608.02781}{http://arxiv.org/abs/1608.02781}.


\bibitem[12]{ambj}
J.~Ambjoern and L.~Chekhov.
\newblock {The matrix model for dessins d’enfants}.
\newblock {\em Ann. Inst. Henri Poincaré D}, 337-361, 2014.
\newblock \burlalt{1404.4240}{http://arxiv.org/abs/1404.4240}.


 

 

\bibitem[13]{Borot:2021}
G.~Borot, S.~Charbonnier, E.~Garcia-Failde F.~Leid and S.~Shadrin.
\newblock {Analytic theory of higher order free cumulants}.
\newblock 2021, \burlalt{2112.12184}{http://arxiv.org/abs/2112.12184}.

 \bibitem[14]{Hock:2022wer}
A.~Hock.
\newblock {On the $x$-$y$ Symmetry of Correlators in Topological Recursion via Loop Insertion Operator}.
\newblock 2022, \burlalt{2201.05357}{http://arxiv.org/abs/2201.05357}.


\bibitem[15]{Eynard:2005}
B.~Eynard and A.~Ferrer.
\newblock {2-matrix versus complex matrix model, integrals over the unitary
  group as triangular integrals}.
\newblock {\em Commun.Math.Phys.}, 264:115--144, 2006,
  \burlalt{hep-th/0502041}{http://arxiv.org/abs/hep-th/0502041}.

\bibitem[16]{Branahl:2020yru}
J.~Branahl, A.~Hock, and R.~Wulkenhaar.
\newblock {Blobbed topological recursion of the quartic Kontsevich model I:
  Loop equations and conjectures},
  \newblock {\em Commun. Math. Phys.} 2022,
\burlalt{2008.12201}{http://arxiv.org/abs/2008.12201}.
\newblock \doi{10.1007/s00220-022-04392-z}.

 \bibitem[17]{Borot:2015hna}
G.~Borot and S.~Shadrin.
\newblock {Blobbed topological recursion: properties and applications}.
\newblock {\em Math. Proc. Cambridge Phil. Soc.}, 162(1):39--87, 2017,
  \burlalt{1502.00981}{http://arxiv.org/abs/1502.00981}.
\newblock \doi{10.1017/S0305004116000323}.


\bibitem[18]{Branahl:2021}
J.~Branahl, A.~Hock.
\newblock {Genus one free energy contribution to the quartic Kontsevich model}.
\newblock 2021, \burlalt{2111.05411}{http://arxiv.org/abs/2111.05411}.

\bibitem[19]{Branahl:2022}
J.~Branahl, A.~Hock.
\newblock {Complete solution of the LSZ Model via Topological Recursion}.
\newblock 2022, \burlalt{2205.12166}{http://arxiv.org/abs/2205.12166}.

  

 
\end{thebibliography}
 
\end{document}